\documentclass[twocolumn,showpacs,pra,aps,superscriptaddress,floatfix]{revtex4}

\usepackage{graphicx}

\begin{document}

\title{Equivalence between the real time Feynman histories and the quantum shutter approaches for the ``passage time" in tunneling}
\author{Gast\'on Garc\'{\i}a-Calder\'on}
\email{gaston@fisica.unam.mx}
\affiliation{Instituto de F\'{\i}sica,
Universidad Nacional Aut\'onoma de M\'exico, 
Apartado Postal {20 364}, 01000 M\'exico, Distrito Federal, M\'exico}
\author{Jorge Villavicencio}
\email{villavics@uabc.mx}
\affiliation{Facultad de Ciencias, Universidad Aut\'onoma de Baja California, 
Apartado Postal 1880, 22800 Ensenada, Baja California, M\'exico}
\author{Norifumi Yamada}
\email{yamada@i1nws1.fuis.fukui-u.ac.jp}
\affiliation{Department of Information Science, Fukui University, 3-9-1 Bunkyo, Fukui, Fukui 910-8507, Japan}
\date{\today}
\begin{abstract}
We show  the equivalence of the functions $G_{\rm p}(t)$ and $|\Psi(d,t)|^2$ for the 
``passage time'' in tunneling. The former, obtained within the framework of the real 
time Feynman histories approach to the tunneling time problem, using the Gell-Mann and 
Hartle's decoherence functional, and the latter involving an exact analytical solution 
to the time-dependent Schr\"{o}dinger equation for cutoff initial waves.
\end{abstract}
%
\maketitle
\section{INTRODUCTION}
The tunneling time problem has remained a controversial issue after the question 
of how long it takes a particle to traverse a classically forbidden region, 
was raised 70 years ago \cite{maccoll}. There are a number of approaches to this 
problem \cite{review}. In this paper, an unexpected close relationship is found 
between a real time Feynman path integral approach \cite{yamadapra, yamada, yamadaun} 
and the quantum shutter approach \cite{gcr97,gcv01}, which are, at first sight, 
unlikely to be related. 

If we use the real time Feynman path integrals \cite{feynman}, we can define the 
^^ ^^ amplitude distribution" of tunneling time as the sum of $e^{iS/\hbar}$ ($S$ 
being the action) over the paths that take a specified amount of time to traverse 
the barrier region. With the amplitude distribution, we can deal with the interesting 
question whether or not a probability distribution is definable for tunneling time \cite{intrinsic}. 
The definability of the probability distribution depends on whether or not the amplitude 
distribution has the property of orthogonality, i.e., whether or not the classes of 
Feynman paths taking time $\tau_1$ and $\tau_2$ ($\tau_1\ne\tau_2$) to traverse the 
region interfere. For rectangular barriers, one of the authors studied the interference 
quantitatively to conclude that (i) a probability distribution is not definable \cite{yamadapra,yamada} 
but (ii) the range of the values of tunneling time is definable \cite{yamada}. 
In Ref. \cite{yamada}, a function $G(\tau)$ is introduced to analyze how different 
classes of Feynman paths (each class being characterized by the value of $\tau$) 
contribute to the tunneling process. The function $G(\tau)$ was 
used to prove the undefinability of the probability distribution and also to estimate 
the range of the tunneling times. For typical opaque barriers, the graph of $G(\tau)$ 
showed a peaked structure near the B\"uttiker-Landauer time \cite{bltime}. It is thus 
clear that $G(\tau)$ is an important quantity for the study of the tunneling time problem. 
It is however important to understand how $G(\tau)$ is 
related to the dynamics of tunneling, which is not evident at all from the Feynman paths 
construction of $G(\tau)$. In the present paper, we will relate $G(\tau)$ (to be 
precise, $G_{\rm p}(\tau)$ as discussed below) to a time-dependent wave function. 
Now, we have to quickly add the following: In general, a Feynman path crosses the 
barrier region many times, so that we can define ^^ ^^ the amount of time taken by a 
Feynman path to traverse the barrier region" in several ways. We can define it as the 
sum of the times during which the Feynman path is within the barrier region \cite{soko}, 
which may be called the resident time of the Feynman path. Or, we can define it as the 
last time the path leaves the barrier region minus the first time it enters the region 
\cite{sch}, which may be called the passage time of the Feynman path. These two different 
definitions at the level of Feynman paths would lead to physically different tunneling 
times, which we shall call the tunneling time of resident time type (\textit{resident time} 
for short) and the tunneling time of passage time type (\textit{passage time} for short). Reference \cite{yamada} concerns the resident time, while Refs. 
\cite{yamadapra, yamadaun} and this paper concern the passage time. We shall attach, 
if necessary, subscript ${\rm r}$ to the quantities for the resident time (e.g., 
$G_{\rm r}(\tau)$) and subscript ${\rm p}$ for the passage time (e.g., $G_{\rm p}(\tau)$). 

Another novel approach, relevant to the tunneling time problem\cite{gcr97}, 
is to consider an analytic time-dependent solution to Schr\"{o}dinger's equation with the initial condition at $t=0$ of an incident cutoff wave, to investigate the time evolution of the probability density through an arbitrary potential barrier. This problem may be visualized as a 
\textit{gedanken experiment} consisting of a shutter, situated at $x=0$, that separates 
a beam of particles from a potential barrier located in the region $0\leq x\leq d$. 
At $t=0$ the shutter is opened, and the probability density rises initially from a 
vanishing value and evolves with time through $x > 0$. At the barrier edge $x=d$, 
the probability density at time $t$, $|\Psi(x,t)|^2$, yields the probability of finding 
the particle after a time $t$ has elapsed. Since initially there is no particle along 
the tunneling region, detecting the particle at the barrier edge at time $t$ should 
provide a relevant time scale of the tunneling process. 
In  recent work, two of the authors \cite{gcv01,gcvarxiv} analyzed the time 
evolution of the probability density $|\Psi(d,t)|^2$ for a rectangular potential 
barrier using the above formalism. 
There, it was found that the probability density at the right barrier edge $x=d$, 
exhibits at short times a transient structure that they named \textit{time-domain resonance}. 
The maximum of the time-domain resonance, occurring at a time $t=t_p$, 
represents the largest probability of finding the particle at $x=d$.  
In Ref. \cite{gcvarxiv} the above authors called 
the attention of the readers to the fact that the shape of the graph of $|\Psi(d,t)|^2$ 
depicted in Fig. 1 of that paper resembles the average shape of the graph in Fig. 2 of 
Ref. \cite{yamada}, which is the graph of $G_{\rm r}(\tau)$. Then they guessed that 
$|\Psi(d,t)|^2$ would be more related to the passage time rather than the resident time. 
In fact, in Ref. \cite{yamadaun}, Yamada has studied $G_{\rm p}(\tau)$ to find that, 
if $\tau$ is simply replaced by $t$, the graph of $G_{\rm p}(\tau)$ for a monochromatic 
case is actually indistinguishable from the graph of $|\Psi(d,t)/T|^2$, where $T$ is 
the transmission amplitude. However, there has been no explicit proof that these two 
functions are really equivalent.

The aim of this paper is to prove that the function $G_{\rm p}(\tau)$ and the probability 
density $|\Psi(d,t)|^2$ under the initial condition stated above are actually related by 
\begin{equation}
\left\vert\frac{\Psi(d,t)}{T}\right\vert^2=G_{\rm p}(t), 
\label{yeq:7}
\end{equation}
thereby establishing a surprising relationship between the two approaches. As a by-product of 
our proof to Eq. (\ref{yeq:7}), we present an alternative derivation, along the transmitted 
region, of the expression for $\Psi(x,t)$  without using the Laplace transform method. This 
derivation is the second purpose of the present paper. 

Section II presents a brief account of the main features of both approaches. Section III 
deals with the proof to Eq. (\ref{yeq:7}) and also with a new derivation of $\Psi(x,t)$. 
In Sec. IV, a numerical example is presented for a rectangular potential barrier in order 
to exhibit the equivalence of both approaches. Concluding remarks are given in Sec.~V. 

\section{THE FORMALISMS}

\subsection{Real Time Feynman path integral approach}

In Ref. \cite{yamada}, Yamada introduced $G(\tau)$ by
\begin{equation}
G(\tau)\equiv\frac{1}{P}\int_0^\tau d\tau_1\int_0^\tau 
d\tau_2\,D[\tau_1;\tau_2]. 
\label{yeq:1}
\end{equation}
In the above expression, $D[\tau_1;\tau_2]$ is the \textit{decoherence functional} for the case of 
tunneling time for transmission and $P$ is the tunneling probability defined by
\begin{equation}
P\equiv\lim_{t\to\infty}\int_d^\infty\!\!dx \vert\Psi(x,t)\vert^2, 
\label{yeq:2}
\end{equation}
where $d$ is the position of the right edge of the barrier. The decoherence functionals were 
formulated in general terms by Gell-Mann and Hartle \cite{gh2} in their version of the consistent 
history approach to quantum mechanics \cite{gh2, gh1, gh3}. The real part of $D[\tau_1;\tau_2]$ 
is a measure of the interference between the classes of Feynman paths that take different amounts 
of time ($\tau_1$ and $\tau_2$) to traverse the barrier region. Roughly speaking, $G(\tau)$ is 
the square modulus of the sum of $e^{iS/\hbar}$ over those paths that take \textit{less than} 
time $\tau$ to traverse the barrier region (to be precise, the result of the sum over paths is 
multiplied by the initial wave function, followed by the integrations over the initial and the 
final positions before and after taking the square, respectively). It is easier to deal with $G(\tau)$ than 
$D[\tau_1,\tau_2]$ since $G(\tau)$ is a real function of one variable, while $D[\tau_1;\tau_2]$ 
is a complex function of two variables. $G(\tau)$ has the following properties: (i) $G(0)=0$ and 
(ii) $G(\infty)=1$. Yamada \cite{yamada} claimed that  (a) if $G(\tau)$ is not an increasing 
function of $\tau$, a probability distribution of tunneling time is not definable, and (b) the 
range $(\tau_<,\tau_>)$ of times is an estimation of the range of values of tunneling time, where 
$\tau_<$ and $\tau_>$ are such that $G(\tau)<\epsilon$ for ${}^\forall \tau < \tau_<$ and 
$|1-G(\tau)|< \epsilon$ for ${}^\forall \tau>\tau_>$, where $0<\epsilon\ll 1$. The first claim 
(a) is based on the \textit{weak decoherence condition} \cite{gh1, gh3} in the consistent history 
approach. 
 
For a particle with wave number $k_0\,(>0)$ impinging on the square barrier of height $V_0$ 
that extends from $x=0$ to $x=d$, $G_{\rm p}$ was found to be \cite{yamadaun} 
\begin{equation}
G_{\rm p}(\tau)=\frac{{k_0}^2}{\pi^2\vert T\vert^2}
\left\vert
\int_{-\infty}^{\infty}\!\! dk \ T(k)\, e^{i k d}\,
\frac{e^{i\hbar ({k_0}^2-k^2)\tau/2m}-1}{k^2-{k_0}^2}
\,\right\vert^2,
\label{yeq:4}
\end{equation}
where $T(k)\equiv T(k,V_0,d)$ is the transmission amplitude for the square barrier when the 
wave number is $k$, and $T= T(k_0)$. 

\subsection{Quantum shutter approach}

A direct access to tunneling phenomena in time domain is to follow the time evolution of the 
Schr\"odinger's wave function. In Refs. \cite{gcvarxiv,gcv01}, two of the authors studied the 
time-dependence of the probability density by using an explicit solution \cite{gcr97} to the 
time-dependent Schr\"{o}dinger equation, with a cutoff plane wave initial condition, 
\begin{eqnarray}
\Psi(x,0)=\cases{e^{ik_0x}-e^{-ik_0x} & for $x<0$\cr 0 & for $x\ge 0$,}
\label{yeq:5}
\end{eqnarray}
impinging on a shutter placed at $x=0$, just at the left edge of the structure that extends 
over the interval $0\leq x\leq d$. The tunneling process begins with the instantaneous opening 
of the shutter at $t=0$, enabling the incoming wave to interact with the potential at $t>0$. 
The exact solution along the transmitted region ($x>d$) reads \cite{gcv01}, 
\begin{eqnarray}
\Psi(x,t)=T(k_0)M(x,k_0;t)-T(-k_0)M(x,-k_0;t)\nonumber\\
-\sum\limits_{n=-\infty }^{\infty}T_{n}M(x,k_n;t).
\label{Psiext}
\end{eqnarray}
In the above expression, the quantities $T(\pm k_0)$ refer to the transmission amplitudes, 
the index $n$ runs over the complex poles $k_n$ of $T(k)$, which are distributed in the 
third and fourth quadrants in the complex $k$ plane, and the factor $T_n$ is defined as 
\begin{equation}
T_n=2ik_0 \frac{u_{n}(0)u_{n}(d)}{k_0^{2}-k_{n}^{2}}{\rm e}^{-ik_n d},
\label{Tn}
\end{equation}
where $\{u_n(x)\}$ are the resonant eigenfunctions \cite{gcr97}, which 
are the solutions to 
\begin{equation}
\frac{d^2 u_n(x)}{dx^2}+\left[{k_n}^2-\frac{2m}{\hbar^2}V(x)\right] u_n(x)=0 
\label{eq:24}
\end{equation}
with outgoing boundary conditions,
\begin{equation}
\left[\frac{d}{dx}u_n(x)\right]_{x=0} = -ik_nu_n(0),
\label{b1}
\end{equation}
and
\begin{equation}
\left[\frac{d}{dx}u_n(x)\right]_{x=d} = ik_nu_n(d).
\label{b2}
\end{equation}

Both the complex poles $\{k_{n}\}$ and the corresponding resonant eigenfunctions $\{u_n(x)\}$ 
can be calculated using a well established method, as discussed elsewhere \cite{gcr97, gcv01}.
Note that from time-reversal considerations \cite{rosenfeld}, the poles $k_{-n}$, seated on the third quadrant of the complex $k$-plane, satisfy $k_{-n}=-k_n^*$ and correspondingly 
$u_{-n}(x)=u_n^*(x)$. In Eq. (\ref{Psiext}), the $M$ functions are defined by 
\begin{eqnarray}
M(x,q;t)&\equiv&\frac{i}{2\pi}\int_{-\infty}^\infty\! dk\, 
\frac{e^{ikx-i\hbar k^2 t/2m}}{k-q}\label{eq:19}\\[2mm]
&=&\frac{1}{2}e^{(imx^2/2\hbar t)}w(iy_q), 
\label{mosh}
\end{eqnarray}
where $q=k_n, \pm k_0$, and $w(iy_q)$ is the complex error 
function \cite{wiz} with the argument $y_q$ given by
\begin{equation}
y_q=e^{-i\pi /4}\sqrt{\frac{m}{2\hbar t}}\left[x-\frac{\hbar q}{m}t\right]. 
\label{yq}
\end{equation}

\section{EQUIVALENCE OF BOTH APPROACHES}

\subsection{Proof of Eq. (\ref{yeq:7})}

We will start from the general relationship between an initial wave function and the time evolved 
wave functions:
\begin{equation}
\Psi(x,t)=\int_{-\infty}^\infty dy\, K(x,t;y,0)\Psi(y,0),
\label{yeq:8}
\end{equation}
where $K(x,t;y,0)$ is the propagator from $(y,0)$ to $(x,t)$. Since our initial wave function is 
vanishing for $x>0$ and since we are interested only in the transmitted region, we need to know 
$K(x,t;y,0)$ only for $y\le 0$ and $x\ge d$, for which it is well-known that
\begin{equation}
K(x,t;y,0)=\int_{-\infty}^\infty \frac{dk}{2\pi}\, T(k) \, 
e^{ik(x-y)-i\hbar k^2 t/2m},
\label{yeq:9}
\end{equation}
which follows from the eigenfunction expansion of the propagator. The initial wave 
function can be expanded as 
\begin{equation}
\Psi(y,0)=\int_{-\infty}^\infty \frac{dk}{\sqrt{2\pi}}\, \phi(k)\, e^{iky},
\label{yeq:10}
\end{equation}
where $\phi(k)$ is the $k$-space wave function. Substituting Eqs. (\ref{yeq:9}) and (\ref{yeq:10}) 
into Eq. (\ref{yeq:8}), we can carry out the integration over $y$ to have
\begin{equation}
\Psi(x,t)=\int_{-\infty}^\infty \frac{dk}{\sqrt{2\pi}}\, \phi(k)T(k)\, 
e^{ikx-i\hbar k^2 t/2m}. 
\label{yeq:11}
\end{equation}
For our initial wave function [Eq. (\ref{yeq:5})],
\begin{eqnarray}
\phi(k)&=&\int_{-\infty}^\infty \frac{dy}{\sqrt{2\pi}} e^{-iky} \Psi(y,0)\nonumber\\
&=&\frac{i}{\sqrt{2\pi}}\left(\frac{1}{k-k_0+i\epsilon}-\frac{1}{k+k_0+i\epsilon}\right),
\label{eq:11}
\end{eqnarray}
where $\epsilon$ is an infinitesimal positive number. Thus, 
\begin{eqnarray}
\Psi(x,t)&=&\frac{i}{2\pi}\int_{-\infty}^\infty \!\!dk
\left\{
\left(\frac{1}{k-k_0+i\epsilon}-\frac{1}{k+k_0+i\epsilon}\right)
\right.
\nonumber\\[2mm]
&&\left.
\phantom{\times\int_{-\infty}^\infty \!\!dk\left\{\left(\right.\right.}
\times T(k)e^{ikx-i\hbar k^2 t/2m}
\right\}
\label{eq:12}
\end{eqnarray}
for $x\ge d$. 

Let us note that, since $\Psi(x,0)=0$ for $x\ge 0$,
\begin{equation}
\int_{-\infty}^\infty dk\,
\left(\frac{1}{k-k_0+i\epsilon}-\frac{1}{k+k_0+i\epsilon}\right)
T(k)\, e^{ikx}=0
\label{eq:13}
\end{equation}
for $x\ge 0$, which is also apparent from the fact that the transmission 
amplitude on the complex $k$-plane has simple poles only in the lower 
half-plane. Owing to Eq.~(\ref{eq:13}), we can rewrite Eq. (\ref{eq:12}) as
\begin{eqnarray}
\Psi(x,t)&=&\frac{i}{2\pi}\int_{-\infty}^\infty \!\!dk
\left\{
\left(\frac{1}{k-k_0+i\epsilon}-\frac{1}{k+k_0+i\epsilon}\right)
\right.
\nonumber\\[2mm]
&&\left.
\phantom{\frac{1}{1}}
\times T(k) e^{ikx}\left(e^{-i\hbar k^2 t/2m}-e^{-i\hbar {k_0}^2 t/2m}\right)
\right\}.\nonumber\\
&& \label{eq:14}
\end{eqnarray}
Apply the following equation in Eq. (\ref{eq:14}). 
\begin{equation}
\frac{1}{k\pm k_0+i\epsilon}={\cal P}\frac{1}{k\pm k_0}-\pi i \delta(k\pm 
k_0).
\label{eq:15}
\end{equation}
We then notice that (i) since $e^{-i\hbar k^2 t/2m}-e^{-i\hbar {k_0}^2 t/2m}=0$
at $k=k_0$, the contributions from the delta functions vanish and (ii) since
$(e^{-i\hbar k^2 t/2m}-e^{-i\hbar {k_0}^2 t/2m})/(k\pm k_0)$ is regular in the 
limit $k\to \mp k_0$, the Cauchy principal value integrals can be replaced by 
the ordinary integrals (i.e., the symbol ${\cal P}$ can be removed). Consequently, 
we have for $x\ge d$ 
\begin{eqnarray}
\Psi(x,t)&=&\frac{i}{2\pi}\int_{-\infty}^{\infty} \!\!dk\,
\left\{
\left(\frac{1}{k-k_0}-\frac{1}{k+k_0}\right)T(k)e^{ikx}
\right.
\nonumber\\ 
&&\left.
\phantom{\int_{-\infty}^{\infty} \!\!dk\,}\times
\left(e^{-i\hbar k^2 t/2m}-e^{-i\hbar {k_0}^2 t/2m}\right)
\right\}
\nonumber\\
&=&\frac{i k_0}{\pi}e^{-i \hbar {k_0}^2t/2m}\nonumber\\
&&\times\int_{-\infty}^{\infty}\!\! dk \ 
T(k)\, e^{i k x}\frac{e^{i\hbar ({k_0}^2-k^2)t/2m}-1}{k^2-{k_0}^2}.
\label{eq:16}
\end{eqnarray}
With this expression for $\Psi$, it is easy to see that $\vert\Psi(d,t)/T\vert^2$ 
agrees with the right-hand side of Eq. (\ref{yeq:4}) if $\tau$ is replaced by $t$. 
This completes the proof of Eq. (\ref{yeq:7}).

\subsection{New derivation of the quantum shutter solution}

As mentioned earlier, the transmission amplitude has in general an infinite number of 
simple poles distributed on the  lower-half of the complex $k$-plane. The transmission 
amplitude may be expanded in terms of its complex poles and corresponding residues by 
using a special form of the Mittag-Leffler theorem due to Cauchy \cite{cauchy}. 
It may be written as \cite{yamadaun}, 
\begin{equation}
T(k)=
\sum_{n=-\infty}^{\infty} \left(\frac{r_n}{k-k_n}+\frac{r_n}{k_n}\right),
\label{eq:17}
\end{equation}
where $r_n$ is the residue of $T(k)$ at $k=k_n$. Using Eq.~(\ref{eq:17}) in 
Eq. (\ref{eq:12}), we have, for $x\ge d$,
\begin{eqnarray}
\Psi(x,t)=&&\frac{i}{2\pi}
\nonumber\\[2mm]
&&
\times\sum_n \int_{-\infty}^\infty \!\!dk
\left\{
\left(\frac{1}{k-k_0+i\epsilon}-\frac{1}{k+k_0+i\epsilon}\right) 
\right.
\nonumber
\\[2mm]
&&\left.\times
\left(\frac{r_n}{k-k_n}+\frac{r_n}{k_n}\right)\,{\rm e}^{ikx-i\hbar k^2 t/2m}
\right\}. 
\label{eq:18}
\end{eqnarray}
If we expand 
\begin{displaymath}
\left (\frac{1}{k-k_0+i\epsilon}-\frac{1}{k+k_0+i\epsilon}\right )
\left (\frac{1}{k-k_n}+\frac{1}{k_n}\right )
\end{displaymath}
and use the partial fraction decompositions, we see that the right-hand side of Eq. (\ref{eq:18}) 
can be expressed as a sum of the integrals of the form of Eq. (\ref{eq:19}). The expansion 
gives four terms, which are $k_n^{-1}(k\pm k_0+i\epsilon)^{-1}$ and 
\begin{eqnarray}
\frac{1}{k\pm k_0+i\epsilon}\,\frac{1}{k-k_n}=\frac{1}{\pm k_0+k_n+i\epsilon}\nonumber\\
\times \left(\frac{1}{k-k_n}-\frac{1}{k\pm k_0+i\epsilon}\right),
\label{eq:20}
\end{eqnarray}
so that Eq. (\ref{eq:18}) becomes, after some algebra,
\begin{eqnarray}
\Psi(x,t)&=&\sum_n\left(\frac{r_n}{k_0-k_n-i\epsilon}+
\frac{r_n}{k_n}\right) M(x,k_0;t)
\nonumber\\
&&-\sum_n\left(\frac{r_n}{-k_0-k_n-i\epsilon}+\frac{r_n}{k_n}
\right)M(x,-k_0;t)\nonumber\\
&&-\sum_n\left(\frac{r_n}{k_0+k_n+i\epsilon}+\frac{r_n}{k_0-k_n-i\epsilon}\right)\nonumber\\
&&\phantom{-}
\times M(x,k_n;t).
\label{eq:21}
\end{eqnarray}
In the limit $\epsilon\to 0$, the sums over $n$ in the first and the second lines on the 
right-hand side of Eq. (\ref{eq:21}) give $T(k_0)$ and $T(-k_0)$, respectively [see Eq. (\ref{eq:17})]. 
We thus obtain
\begin{eqnarray}
\Psi(x,t)=T(k_0)M(x,k_0;t)-T(-k_0)M(x,-k_0;t)\nonumber\\
-2k_0\sum_{n=-\infty}^\infty\frac{r_n}{{k_0}^2-{k_n}^2}\, M(x,k_n;t).
\label{eq:22}
\end{eqnarray}

Our goal here is to derive Eq. (\ref{Psiext}). In fact, Eqs. (\ref{eq:22}) and 
(\ref{Psiext}) are the same because of the relationship
\begin{equation}
r_n=iu_n(0)u_n(d)e^{-ik_n d}.
\label{eq:25}
\end{equation}
We shall prove Eq. (\ref{eq:25}) to conclude this section. It is helpful to 
consider the outgoing Green's function $G^+(x,x';k)$, which is the solution to
\begin{equation}
\frac{\partial^2 G^+(x,x';k)}{\partial x^2}+\left[{k_n}^2-\frac{2m}{\hbar^2}V(x)\right] G^+(x,x';k)=\delta(x-x')
\label{eq:26}
\end{equation}
with the outgoing boundary conditions. We first use the fact that
$G^+(x,x';k)$ can be written in terms of the resonant states as \cite{expansion}, 
\begin{equation}
G^+(x,x';k)=\sum_{n=-\infty}^{\infty}\frac{u_n(x)u_n(x')}{2k_n(k-k_n)} \quad (0\le x, x'\le d).
\label{eq:27}
\end{equation}
The above expansion holds provided that the resonant eigenfunctions $u_n(x)$ are normalized according 
to the condition \cite{gcr97} 
\begin{equation}
\int_0^du_n^2(x)dx + i \frac {u_n^2(0)+u_n^2(d)}{2k_n}=1. 
\label{norm}
\end{equation}
Next we use the fact that the transmission amplitude and $G^+(0,d;k)$ are related by \cite{gc87}
\begin{equation}
T(k)=2ik G^+(0,d;k) e^{-ikd}.
\label{eq:28}
\end{equation}
From Eq. (\ref{eq:27}), we have 
\begin{equation}
\lim_{k\to k_n} (k-k_n)G^+(0,d;k)=u_n(0)u_n(d)/2k_n,
\label{28a}
\end{equation}
while from Eq. (\ref{eq:28}) together with Eq. (\ref{eq:17}), we have
\begin{equation}
\lim_{k\to k_n} (k-k_n)G^+(0,d;k)=r_n e^{ik_n d}/2ik_n.
\label{28b}
\end{equation}
Equating the two results, we obtain Eq. (\ref{eq:25}).

\section{Example}

\begin{figure}[!tbp]
\rotatebox{0}{\includegraphics[width=3.6in]{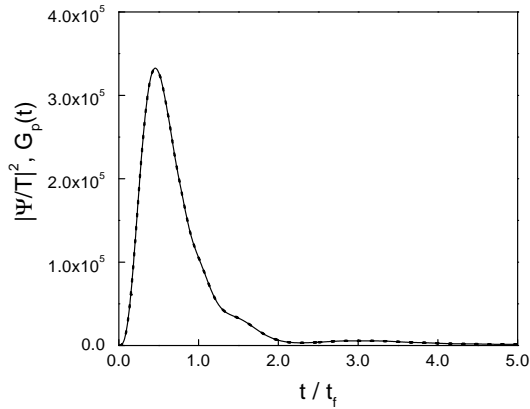}}
\caption{Plot of $|\Psi(d,t)|^2$ (solid line) and
$G_{\rm p}(t)$ (dotted line) at the barrier edge $x=d=10.083$ nm
as a function of time in units of the free passage time $t_f$.
The time-dependent solution is normalized to the transmission 
coefficient $|T|^2$.}
\label{fig1}
\end{figure}

To exemplify the time evolution of the probability density we consider the set of 
parameters: $V_0=0.70$ eV, $d=10.083$ nm, $E=0.140$ eV, $m=0.067\,m_e$ ($m_e$ being 
the bare electron mass), inspired in semiconductor quantum structures \cite{qs}. 
In this particular example, the potential barrier parameters are chosen in such 
a way that $k_0d=V_0/E=5$, where $k_0=[2mE]^{1/2}/\hbar$. The opacity 
$\alpha$ of the barrier is defined as $\alpha=k'd$, where $k'=[2mV_0]^{1/2}/\hbar$. 
In our case $\alpha=11.18$, corresponding to an opaque barrier ($\alpha \gg 1$). 
The solid line in Fig. \ref{fig1} shows $|\Psi(d,t)|^2$ calculated with 
Eq. (\ref{Psiext}) at the barrier edge $x=d$ as a function of time in units of 
the free passage time $\tau_f=md/\hbar k_0=11.753$ fs. 
At early times one sees a {\it time-domain resonance} structure \cite{gcv01}. The 
maximum of this transient structure represents the largest probability to find the 
tunneling particle at the barrier edge $x=d$. In our example, as shown in Fig. \ref{fig1}, 
the maximum of the {\it time-domain resonance} occurs at $t_p=5.347$ fs, faster 
than the free passage time across the same distance of $10.083$ nm, that is, 
$t_p/t_f=0.455$. From $t/t_f=2.0$ onward the probability density approaches 
essentially to its asymptotic value. We have also included in Fig. \ref{fig1} 
the plot of $G_{\rm p}(t)$ (dotted line) calculated from Eq. (\ref{yeq:4}) for 
the same set of parameters; it is indistinguishable from the previous calculation, 
i.e., both curves coincide exactly. 

\section{Concluding remarks}

We have found a surprising relationship between the real time Feynman histories approach 
and an analytical expression for the probability density for cutoff initial waves involving 
the quantum shutter setup for the ``passage time'' in tunneling. This may prove to be of 
interest in the pursue of elucidating the notion of tunneling time through a classically 
forbidden region.

\begin{acknowledgments}

G. G-C. and J.V. acknowledge partial financial support of DGAPA-UNAM under grant No. IN101301. 
N.~Y. is grateful to the Instituto de F\'{\i}sica, UNAM for their hospitality and for financial 
support from the Tom\'as Brody Spitz Chair. Also, N. Y. acknowledges partial financial support 
of Grant-in-Aids for Scientific Research from the Ministry of Education, Culture, Sports, 
Science and Technology, Japan and  thanks Professor H. Yamamoto and Professor S. Takagi 
for valuable discussions and to the Information Synergy Center at Tohoku University 
for CPU time. 
\end{acknowledgments}


\begin{thebibliography}{22}
\bibitem{intrinsic} Here we consider such a probability distribution of tunneling 
times that is intrinsic to the scattering event.
\end{thebibliography}
\end{document}